# Fluctuation and Inertia

**Liangsuo Shu[1,2], Xiaokang Liu[2], Kaifeng Cui[3,4], Zhichun Liu[2], Wei Liu[2*]**

[1]School of Physics, Huazhong University of Science & Technology. Wuhan, China.
[2]School of Energy and Power Engineering, Huazhong University of Science & Technology. Wuhan, China.
[3]University of Chinese Academy of Sciences, Beijing, China
[4]Key Laboratory of Atom Frequency Standards, Wuhan Institute of Physics and Mathematics, Chinese Academy of Sciences, Wuhan, China

## Abstract

In this work, using Jacobson's idea: "$\delta Q=TdS$ hold for all the local Rindler causal horizons through each spacetime point", we found that the transitions between the excited and ground state of a particle in a linear acceleration satisfy the fluctuation theorem. The up transition from the ground state to the excited state is an entropy-decreasing process, which requires an external force to contribute equal entropy increase to satisfy the second law of thermodynamics.

L. S. : liangsuo_shu@hust.edu.cn
X. L. : xk_liu@hust.edu.cn
K. C. : cuikaifeng@wipm.ac.cn
Z. L. : zcliu@hust.edu.cn
W. L.: w_liu@hust.edu.cn

## 1. Introduction

There should be no doubt that the principle of inertia can be among the most important laws of physics. Look backing at the history of physics, we can find that the deepening understanding of inertia has play important roles in both the development of classic mechanics and the discovery of general relativity. Study on the details of the gravitational coupling between matter and spacetime which results in inertia may provide more useful information. According to the researches in the references which will be mentioned below, thermodynamics is proved to be an effective method.

In 1970s, the discovery of the Hawking radiation have shown deep relations among thermodynamics, relativity, and quantum mechanics [1–3]. From the end of last century, there are new important developments in this area. On the one hand, the close relationships between the second law of thermodynamics and some basic assumptions in both relativity and quantum mechanics were discovered [4–7]. On the other hand, the attempt to interpret the field equations of gravitational theories in a thermodynamic context is one important topic of the theoretical progress in gravitational interaction at the same time [8–10].

Bell and Leinaas [11] had got the up and down transition probabilities between the ground state and the excited state of a uniformly accelerating particle. Using Jacobson's idea in his thermodynamic derivation of the Einstein equation [12], we found the transitions satisfy the fluctuation theorem. The thermodynamic role of external force is to contribute equal entropy increase to satisfy the second law of thermodynamics. If the external force is absent, the particle will tend to return to the ground state under the effect of the entropy increase principle.

## 2. Fluctuation theorem in a linear acceleration

According to the work of Bell and Leinaas [11], the up and down transition probabilities between the ground state ($E_0$) and the excited state ($E=E_0+\Delta E$) of a uniformly accelerating particle satisfy

$$\frac{P_+}{P_-} = \exp(-\frac{2\pi c \Delta E}{\hbar a}) \quad (1)$$

where $a$ is the acceleration, $P_+$ and $P_-$ are the up and down probabilities, respectively. The Unruh temperature [13] experienced by the uniformly accelerating particle is

$$T = \frac{\hbar a}{2\pi c k_B} \quad (2)$$

Substituting equation (2) into (1), we can get

$$\frac{P_+}{P_-} = \exp(-\frac{\Delta E}{k_B T}) \quad (3)$$

When Jacobson [12] deduced the Einstein equation using thermodynamics method, the fundamental relation

$$\delta Q = TdS \quad (4)$$

is assumed to hold for all the local Rindler causal horizons through each spacetime point, with $\delta Q$ and $T$ are the energy flux and the Unruh temperature seen by an accelerated observer just inside the horizon. In other word, the energy flux across a causal horizon is a kind of heat flow in spacetime dynamics. For an isothermal process, integrating equation (4) gives

$$\Delta Q = T\Delta S \quad (4.a)$$

During an accelerating process, the energy flux which leads to the energy change of the accelerating particle, $\Delta E$, has to cross its local Rindler causal horizon and therefore be a kind of heat flow in spacetime dynamics [12]. According to the conservation of energy, the heat extracted from the space, $\Delta Q$, when the particle harvest energy of $\Delta E$ can be written as

$$\Delta Q = -\Delta E \quad (5)$$

From equations (4.a) and (5), the change in the entropy of space, $\Delta S$, can be written as

$$\Delta S = \frac{\Delta Q}{T} = \frac{-\Delta E}{T} \quad (6)$$

From equation (6), we can find that, during an accelerating process, the down transition from $E$ to $E_0$ is a spontaneous entropy-increasing process, while the up transition from $E_0$ to $E$ is an entropy-decreasing process. Substituting equation (6) into equation (3) gives

$$\frac{P_+}{P_-} = \exp(\frac{\Delta S}{k_B}) \quad (7)$$

This is in fact the fluctuation theorem for a non-equilibrium system [14]. Therefore, the up transition is an entropy-decreasing fluctuation. In order to occupy the excited state to complete the acceleration, there must be an external intervention to contribute additional entropy increase to satisfy the second law of thermodynamics. Otherwise, the particle, which occupies excited state temporarily after the fluctuation, will spontaneously return to the ground state because of the principle of entropy increase. In an accelerating process, the external intervention is an external force acting on the particle.

Jacobson's idea [12] means that the space works as a heat source which transfer the work done by the external force to heat that crosses the local Rindler causal horizon of an accelerating particle. In the quantum picture of space [15] which can be regarded as composite of gravitons, this is not a problem.

3. **Inertia force as entropic force**

In a uniformly acceleration process, the particle trends to be in the ground state under the entropy increase principle. As a measure of this tendency, the inertia force is therefore an entropic force.

Assuming the proper time interval, needed by the uniformly accelerating particle to harvest $dE$ is $d\tau$, in the instantaneous rest frame of the particle,

$$dl = cd\tau \tag{8}$$

where $dl$ is the proper difference of the particle in 4-D spacetime during the proper time difference of $d\tau$. The entropic inertia force $F_i$ will be

$$F_i dl = TdS \tag{9}$$

Rearranging equation (9) and substituting equation (5) into it, we can obtain

$$F_i = -\frac{dE}{dl} = -\frac{dE}{cd\tau} = -\frac{\gamma \mathbf{v}}{c}\frac{d\mathbf{p}}{dt} \tag{10}$$

where $\mathbf{v}$, $\mathbf{p}$ are the 3-space vectors describing the velocity and the momentum of the particle respectively.

For the particle applied by the inertia force, $F_i$ is in the time dimension of its instantaneous rest frame: its direction is opposite to the direction of time. To complete a real acceleration process with an acceleration of $a$, the application of an external force is necessary to balance the role of inertial forces to ensure that the total entropy change of the acceleration process is non-negative. Beside inertia force and gravity, other external forces are four-force. The time-dimensional component of the external four-force should equal to the the inertia force.

$$F_i + F_t = 0 \tag{11}$$

where $F_t$ is the time-dimensional component of the external four-force. From equations (10) and (11), we can obtain

$$F_t = -F_i = \frac{\gamma \mathbf{v}}{c}\frac{d\mathbf{p}}{dt} = \frac{\gamma \mathbf{v}}{c}\mathbf{f} \tag{12}$$

where $\mathbf{f}$ is the 3-space force vector acting on the particle in acceleration. Taking into account the combined effect of the external force and the inertia force, a real acceleration process is an isentropic process. The inertial force is the result of the tendencye of the entropy of the local space to reach the extreme value.

## 4. Conclusion and discussion

Based on the works of Jacobson [12], we discussed the relationship between the inertia principle and the second law of thermodynamics. The inertia principle is found to a result of the second law of thermodynamics. A particle in inertial motion is a state with maximum entropy.

As a statistical law, the probability that the second law will be violated is not zero. Therefore, the fluctuation is unavoidable, especially when $\Delta S$ is small enough. However, according to the fluctuation theorem, the down transition probability from the excited state to the ground state is bigger than the probability of the reverse process, the effect of the fluctuation is limited when the time scale is large enough.

### Acknowledgements

This works is supported by the National Science Foundation of China (No. 51736004 and No.51776079).